\documentclass[12pt]{article}
\font\mybb=msbm10 at 12pt
\def\bb#1{\hbox{\mybb#1}}
\def\ZZ {\bb{Z}}
\def\RR {\bb{R}}

\usepackage{fullpage,epsfig,fancyheadings}

\usepackage[psamsfonts]{euscript}

\usepackage{epic}

\usepackage{amstex}

\newcommand{\beq}{\begin{equation}}
\newcommand{\eeq}{\end{equation}}

\begin{document}
\vspace*{-.6in}
\thispagestyle{empty}
\begin{flushright}
RU-95-68\\
CALT-68-2025\\
hep-th/9510086
\end{flushright}
\baselineskip = 20pt

\vspace{.5in}
{\Large
\begin{center}
{\bf The Power of M Theory}
\end{center}}
\vspace{.4in}

\begin{center}
John H. Schwarz\\
\emph{Rutgers University, Piscataway, NJ 08855-0849 USA\\
and\\
California Institute of Technology, Pasadena, CA  91125
USA\footnote{Permanent address.}}
\end{center}
\vspace{1in}

\begin{center}
\textbf{Abstract}
\end{center}
\begin{quotation}
\noindent A proposed duality between type IIB superstring theory
on $\RR^9 \times S^1$ and a conjectured 11D fundamental theory
(``M theory'') on $\RR^9 \times T^2$ is investigated.
Simple heuristic reasoning
leads to a consistent picture relating the various $p$-branes
and their tensions in each theory. Identifying the M theory on
$\RR^{10} \times S^1$ with type IIA superstring theory
on $\RR^{10}$, in a similar fashion, leads to various
relations among the $p$-branes of the IIA theory.
\end{quotation}
\vfil

\newpage

\pagenumbering{arabic} 

\section{Introduction}

Recent results indicate that if one assumes the existence of a
fundamental theory in eleven dimensions
(let's call it the `M theory'\footnote{This name was suggested
by E. Witten.}), this provides
a powerful heuristic basis for understanding various results in string theory.
For example, type II superstrings can be understood as arising from
a supermembrane in eleven dimensions \cite{bergshoeff}
by wrapping one dimension of a toroidal supermembrane on a circle
of the spatial geometry \cite{duffa,duffstelle,townsend11,witten}.
Similarly, when the spatial geometry
contains a $K3$, one can obtain a heterotic string by wrapping
a five-brane with the topology of $K3 \times S^1$ on the
$K3$ \cite{harvey,townsend7}.
This provides a very simple heuristic for understanding `string-string
duality' between type IIA  and heterotic strings in
six dimensions \cite{hulla,duffsix,sensix,harvey,vafawitten,duffliu}.
One simply considers the M theory
on $\RR^6 \times S^1 \times K3$. This obviously contains both
type II strings and heterotic strings, arising by the two wrappings
just described. Moreover, since the membrane and 5-brane are
electric-magnetic duals in 11 dimensions, the two strings are dual in
six dimensions, and so it is natural that the strong-coupling
expansion of one corresponds to the weak-coupling expansion of the
other. The remarkable thing about this kind of reasoning is that
it works even though
we don't understand how to formulate the M theory as a quantum theory.
It is tempting to say that
the success of the heuristic arguments that have been given
previously, and those that will be given here, suggest
that there really is a well-defined quantum M theory even when
perturbative analysis is not applicable. The only thing that
now appears to be special about strings is the possibility
of defining a perturbation expansion. In other respects, all
$p$-branes seem to be more or less equal \cite{townsendb,becker}.

Recently, I have analyzed heuristic relationships between Type II
strings and the M theory \cite{twob}. The approach was to compare the 9D
spectrum of the M theory on $\RR^9\times T^2$ with
the IIB theory on $\RR^9\times S^1$. A nice correspondence
was obtained between states arising from the supermembrane
of the M theory and the strings of the IIB theory. The purpose of
this paper is to extend the analysis to include higher $p$-branes
of both theories, and to see what can be learned from imposing
the natural identifications.

Let us begin by briefly recalling the results obtained in
\cite{twob}. We compared the M theory compactified on a
torus of area $A_{M}$ in the
canonical 11D metric $g^{(M)}$ with the IIB theory
compactified on a circle of radius $R_B$
(and circumference $L_B = 2 \pi R_B$) in the
canonical 10D IIB metric $g^{(B)}$. The canonical IIB metric is the
convenient choice, because it is invariant under the $SL(2,\RR)$
group of IIB supergravity. By matching the 9D spectra of
the two models (especially for BPS saturated states),
the modular parameter $\tau$
of the torus was identified with the modulus $\lambda_0
= \chi_0 + i e^{-\phi_0}$, which is
the vev of the complex scalar field of the IIB theory. This
identification supports the conjectured non-perturbative $SL(2,\ZZ)$
duality symmetry of the IIB theory. (This was also noted by Aspinwall
\cite{aspinwall}.)

A second result was that the IIB theory has an infinite
spectrum of strings, which forms an $SL(2,\ZZ)$ multiplet.
The strings, labelled
by a pair of relatively prime integers $(q_1,q_2)$, were
constructed as solutions of the low-energy 10D IIB supergravity
theory using results in Refs. \cite{dabholkara,hullb}. They have
an $SL(2,\ZZ)$ covariant spectrum of tensions given by
\begin{equation}
T_{1q}^{(B)}  =  \Delta_q^{1/2} T_1^{(B)}, \label{t1qb}
\end{equation}
where $T_1^{(B)}$ is a constant with dimensions of mass-squared,
which defines the scale of the theory, and\footnote {Equation
(\ref{Delta}) was
given incorrectly in the original versions of my previous papers
\cite{twob}. Also, $T_1^{(B)}$ and $T_2^{(M)}$ were called $T$ and
$T_{11}$, and $A_M$ was called $A_{11}$.
A more systematic notation is now desirable.}
\begin{equation}
\Delta_q  = e^{\phi_{0}} (q_1 - q_2 \chi_0 )^2 +
e^{-\phi_{0}} q_2^2.\label{Delta}
\end{equation}
Note that strings with $q_2 \neq 0$, those carrying RR charge,
have tensions that, for small string coupling $g_B = e^{\phi_0}$,
scale as $g_B^{-1/2}$. The usual $(1,0)$ string, on the other hand,
has $T \sim g_B^{1/2}$. In the string metric, these become
$g_B^{-1}$ and $1$, respectively.

The mass spectrum of point particles (zero-branes)
in nine dimensions obtained from the two
different viewpoints were brought into agreement (for BPS saturated
states, in particular) by identifying winding modes
of the family of type IIB strings on the circle with Kaluza--Klein modes
of the torus and by identifying Kaluza-Klein modes of the circle with
wrappings of the supermembrane (2-brane) on the torus.\footnote{The
rule that gave sensible results was
to allow the membrane to cover the torus any number of times
(counting orientation), and to identify all the different ways of doing
as equivalent.  For other problems (such as Strominger's
conifold transitions \cite{strominger}) a different rule is required.
As yet, a
single principle that gives the correct rule for all such problems
is not known. I am grateful to A. Strominger
and A. Sen for correspondence concerning this issue.}
The 2-brane of the M theory has a tension (mass per unit area)
in the 11D metric denoted $T_2^{(M)}$.
If one introduces a parameter $\beta$
to relate the two metrics ($g^{(B)} = \beta^2 g^{(M)}$), then
one finds the following relations \cite{twob}
\begin{equation}
(T_1^{(B)} L_B^2)^{-1} = {1 \over (2 \pi)^2}
T_2^{(M)} A_{M}^{3/2}, \label{rbeqn}
\end{equation}
\begin{equation}
\beta^2 = A_{M}^{1/2} T_2^{(M)}/T_1^{(B)}. \label{betaeqn}
\end{equation}
Both sides of eq. (\ref{rbeqn}) are dimensionless numbers,
which are metric independent, characterizing the
size of the compact spaces. Note that, since $T_1^{(B)}$ and
$T_2^{(M)}$ are fixed constants, eq. (\ref{rbeqn}) implies that
$L_B \sim A_{M}^{-3/4}$.

Strings (1-branes) in nine dimensions were also matched. A toroidal
2-brane with one of its cycles wrapped on the spatial two-torus
was identified
with a type IIB string. When the wrapped cycle of the 2-brane is mapped to
the $(q_1, q_2)$ homology class of the spatial torus and taken to
have minimal length $L_q = (A_M/\tau_2)^{1/2} |q_2\tau -q_1| =
(A_M \Delta_q)^{1/2}$,
this gives a spectrum of string tensions
in the 11D metric $T_{1q}^{(M)} = L_q T_2^{(M)}$. Converting to the
IIB metric by $T_{1q}^{(B)} = \beta^{-2}T_{1q}^{(M)} $ precisely
reproduces the previous formula for $T_{1q}^{(B)}$
in eq. (\ref{t1qb}), which therefore
supports the proposed interpretation.

\section{More Consequences of M/IIB Duality}

Having matched 9D point particles (0-branes) and strings (1-branes)
obtained from the IIB and M theory pictures,
let us now explore what additional information
can be obtained by also matching $p$-branes with $p = 2,3,4,5$
in nine dimensions.\footnote{For useful background on $p$-branes see
Refs. \cite{horowitz,townsend92,duffreview}.}
It should be emphasized that even though
we use extremely simple classical reasoning, it ought to be
precise (assuming the existence of an M theory), because we
only consider $p$-branes whose tensions are related to their
charges by saturation of a BPS bound. This means that the relations
that are obtained
should not receive perturbative or non-perturbative quantum corrections.
This assumes that the supersymmetry remains unbroken, which is certainly
believed to be the case.

We begin with $p=2$. In the M theory
the 2-brane in 9D is the same one as in 11D. In the IIB
description it is obtained by wrapping an $S^1$ factor
in the topology of a self-dual
3-brane once around the spatial circle. Denoting the 3-brane
tension by $T_3^{(B)}$, its wrapping gives a 2-brane with tension
$L_B T_3^{(B)}$. Converting to the 11D metric and identifying
the two 2-branes gives the relation
\begin{equation}
T_2^{(M)} = \beta^3 L_B T_3^{(B)}.
\end{equation}
Using eqs. (\ref{rbeqn}) and (\ref{betaeqn}) to eliminate $L_B$ and $\beta$
leaves the relation
\begin{equation}
T_3^{(B)} = {1 \over 2\pi}\Big( T_1^{(B)} \Big)^2 . \label{threeb}
\end{equation}
The remarkable thing about this result is that it is a relation
that pertains entirely to the IIB theory, even though it was deduced
from a comparison of the IIB theory and the M theory. It should also
be noted that the tension $T_3^{(B)}$ is independent of the string
coupling constant, which implies that in the string metric it
scales as $g_B^{-1}$.

Next we consider 3-branes in nine dimensions. The only way they can
arise in the M theory is from wrapping a 5-brane of suitable topology
(once) on the spatial torus. In the IIB theory the only 3-brane is the
one already present in ten dimensions. Identifying the tensions of these
two 3-branes gives the relation
\begin{equation}
T_5^{(M)} A_{M} = \beta^4 T_3^{(B)}.
\end{equation}
Eliminating $\beta$ and substituting eq. (\ref{threeb}) gives
\begin{equation}
T_5^{(M)} = {1 \over 2 \pi}\Big( T_2^{(M)}\Big)^2 . \label{t5m}
\end{equation}
This result pertains entirely to the M theory. Section 3
of ref. \cite{duffliu} analyzed
the implication of the Dirac quantization rule \cite{nepomechie}
for the charges of
the 2-brane and 5-brane in the M theory. It was
concluded that (in my notation) $\pi T_5^{(M)}/ (T_2^{(M)})^2$
should be an integer. The present analysis says that it is $1/2$. Indeed,
I believe that eq. (\ref{t5m}) corresponds to the minimum product
of electric and magnetic charges allowed by the quantization
condition. It is amusing that simple classical
reasoning leads to a non-trivial quantum result.

Next we compare 4-branes in nine dimensions. The IIB theory
has an infinite $SL(2,\ZZ)$ family of 5-branes. These are labeled by
a pair of relatively prime integers $(q_1,q_2)$, just as the IIB
strings are. The reason is that they carry a pair of magnetic charges
that are dual to the pair of electric charges carried by the strings.
Let us denote the tensions of these 5-branes in the IIB metric by
$T_{5q}^{(B)}$. Wrapping each of them once around the spatial circle
gives a family of 4-branes in nine dimensions with tensions
$L_B T_{5q}^{(B)}$. In the M theory we can obtain 4-branes
in nine dimensions by considering 5-branes with an $S^1$ factor
in their topology and mapping the $S^1$ to a $(q_1,q_2)$ cycle of the
spatial torus. Just as when we wrapped the 2-brane this way, we assume
that the cycle is as short as possible, {\it i.e.,} its length is $L_q$.
Identifying the two families of 4-branes obtained in this way gives
the relation
\begin{equation}
L_q T_5^{(M)} = \beta^5 L_B T_{5q}^{(B)}.
\end{equation}
Substituting the relations \cite{twob}
\begin{equation}
L_q = \Delta_q^{1/2} T_1^{(B)} \beta^2 / T_2^{(M)}
\end{equation}
and
\begin{equation}
L_B \beta^3 = 2 \pi T_2^{(M)} / \big(T_1^{(B)}\big)^2 \label{rbthree}
\end{equation}
and using eq. (\ref{t5m}) gives
\begin{equation}
T_{5q}^{(B)} = {1 \over (2 \pi )^2} \Delta_q^{1/2}
\Big( T_1^{(B)}\Big)^3. \label{fiveqb}
\end{equation}
This relation pertains entirely to the IIB theory. Since 5-brane charges
are dual to 1-brane charges, they transform contragrediently under
$SL(2,\RR)$. This means that {\it in this case} $q_1$ is a
magnetic R-R charge
and $q_2$ is a
magnetic NS-NS charge. Thus 5-branes with pure R-R charge have a
tension that scales as $g_B^{1/2}$ and ones with any NS-NS charge
have tensions that scale as $g_B^{-1/2}$. Converting to the string metric,
these give $g_B^{-1}$ and $g_B^{-2}$, respectively. Of course, $g_B^{-2}$
is the characteristic behavior
of ordinary solitons, whereas $g_B^{-1}$ is the
remarkable intermediate behavior that is characteristic of all $p$-branes
carrying R-R charge. It is gratifying that these expected properties
emerge from matching M theory and IIB theory $p$-branes.

We have now related all 1-brane, 3-brane, and 5-brane
tensions of the IIB theory in ten dimensions, so that they
are determined by a single scale.
We have also related the 2-brane and 5-brane tensions of the M theory
in eleven dimensions, so they are also given by a single scale.
The two sets of
scales can only be related to one another after compactification, however,
as the only meaningful comparison is provided by
eqs. (\ref{rbeqn}) and (\ref{betaeqn}).

All that remains to complete this part of the story, is to compare
5-branes in nine dimensions. Here something a little different happens.
As is well-known, compactification
on a space $K$ with isometries (such
as we are considering), so that the complete manifold is $K \times \RR^d$,
give rise to massless vectors in $d$
dimensions. Electric charges that couple to
these vectors correspond to Kaluza--Klein
momenta and are carried by point-like 0-branes. The dual magnetic
objects are $(d-4)$-branes. This mechanism therefore contributes
``Kaluza--Klein 5-branes'' in nine dimensions. However, which
5-branes are the Kaluza--Klein ones depends
on whether we consider the M theory
or the IIB theory. The original 5-brane of the M theory corresponds
to the unique Kaluza--Klein 5-brane of the IIB theory, and the
$SL(2,\ZZ)$ family
of 5-branes of the IIB theory corresponds to the Kaluza--Klein
5-branes of the M theory. The point is that there are three vector
fields in nine dimensions which transform as a singlet and a doublet
of the $SL(2,\RR)$ group. The singlet arises \`a la Kaluza--Klein
in the IIB theory and from the three-form gauge field in the M
theory. Similarly, the doublet arises from the doublet of two-form
gauge fields in the IIB theory and \`a la Kaluza--Klein in the M theory.

We can now use the identifications described above to deduce
the tensions of Kaluza--Klein 5-branes in nine dimensions. The KK
5-brane of the IIB theory is identified with the fundamental
5-brane of the M theory, which implies that its tension is
$T_5^{(B)} = \beta^{-6} T_5^{(M)}$. Combining this with
eq. ~(\ref{rbthree}) gives
\begin{equation}
T_5^{(B)} = {1 \over (2\pi)^3} L_B^2 \big(T_1^{(B)}\big)^4 .
\end{equation}
Note that this diverges as $L_B \to \infty$, as is expected
for a Kaluza--Klein magnetic $p$-brane. Similarly the $SL(2,\ZZ)$
multiplet of KK 5-branes obtained from the M theory must have
tensions that match the 5-branes of the 10D IIB theory. This implies
that $T_{5q}^{(M)} = \beta^6 T_{5q}^{(B)}$. Substituting
eqs.~(\ref{betaeqn}) and (\ref{fiveqb}) gives
\begin{equation}
T_{5q}^{(M)} = {1 \over (2\pi)^2} A_{M}^{3/2} \big(T_2^{(M)}\big)^3
\Delta_q^{1/2} . \label{fiveq}
\end{equation}
This also diverges as $A_{M} \to \infty$, as is expected.
As a final comment, we note that if all tensions are rescaled
by a factor of $2 \pi$ (in other words, equations are rewritten
in terms of $\tilde T = T/2 \pi$), then all the relations we have
obtained in eqs. (\ref{rbeqn}) -- (\ref{fiveq})
have a numerical coefficient of unity.

\section{The IIA Theory}

The analysis given above is easily extended to the IIA theory
in ten dimensions. The IIA theory is simply interpreted
\cite{townsend11,witten} as the M theory on $\RR^{10} \times S^1$.
Let $L=2\pi r$ be the circumference of the circle in the 11D
metric $g^{(M)}$. The string metric of the IIA theory is given
by $g^{(A)} = {\rm exp}(2\phi_A/3)g^{(M)}$, where $\phi_A$ is
the dilaton of the IIA theory. The IIA string coupling constant $g_A$
is given by the vev of ${\rm exp}\, \phi_A $. These facts immediately
allow us to deduce the tensions $T_p^{(A)}$ of IIA $p$-branes for $p=1,2,4,5$.
The results are
\begin{equation}
T_1^{(A)} = g_A^{-2/3} L T_2^{(M)}, \label{t1a}
\end{equation}
\begin{equation}
T_2^{(A)} = g_A^{-1}  T_2^{(M)}, \label{t2a}
\end{equation}
\begin{equation}
T_4^{(A)} = g_A^{-5/3} L T_5^{(M)}, \label{t4a}
\end{equation}
\begin{equation}
T_5^{(A)} = g_A^{-2} T_5^{(M)}. \label{t5a}
\end{equation}
Since $T_1^{(A)}$ and $T_2^{(M)}$ are constants, eq. (\ref{t1a})
gives the scaling rule $g_A \sim L^{3/2}$ \cite{witten,twob}.
Substituting eqs. (\ref{t1a}) and (\ref{t5m}) into eqs. (\ref{t4a})
and (\ref{t5a}) gives
\begin{equation}
T_4^{(A)} = {1 \over 2 \pi} g_A^{-1} T_1^{(A)} T_2^{(M)}
= {1 \over 2 \pi} T_1^{(A)} T_2^{(A)}, \label{t4anew}
\end{equation}
\begin{equation}
T_5^{(A)} = {1 \over 2 \pi} \big(T_2^{(A)}\big)^2. \label{t5anew}
\end{equation}
Again we have found the expected scaling behaviors: $g_A^{-1}$ for the
2-brane and 4-brane, which carry R-R charge, and $g_A^{-2}$ for
the NS-NS solitonic 5-brane. Combining eqs. (\ref{t4anew}) and
(\ref{t5anew}) gives
\begin{equation}
T_2^{(A)}T_4^{(A)} = T_1^{(A)} T_5^{(A)}. \label{t2t4}
\end{equation}
This shows that the quantization condition for the corresponding charges
is satisfied with the same (minimal) value in each case.

The IIA theory also contains an infinite spectrum of BPS saturated
0-branes (aka `black holes') and a dual 6-brane, which are
of Kaluza--Klein origin like those
discussed earlier in nine dimensions. Since the Kaluza--Klein vector
field is in the R-R sector,
the tensions of these should be proportional to $g_A^{-1}$, as was
demonstrated for the 0-branes in \cite{witten}.

\section{P-Branes With $P \geq 7$}

The IIB theory has a 7-brane, which
carries magnetic $\chi$ charge. The way to understand this
is that $\chi$ transforms under $SL(2,\RR)$ just like the axion
in the 4D N=4 theory. It has a Peccei-Quinn translational symmetry
(broken to discrete shifts by quantum effects), which means
that it is a 0-form gauge field. As a consequence,
the theory can be recast in terms of a dual 8-form potential.
Whether or not one does that, the classical supergravity equations
have a 7-brane solution, which is covered by the general analysis of
\cite{horowitz}, though that paper only considered $p \leq 6$.
Thus the 7-brane in ten dimensions is analogous to a string
in four dimensions. Let us call the tension of the IIB 7-brane $T_7^{(B)}$.

The existence of the 7-brane in the 10D IIB theory suggests that
after compactification on a circle, the resulting 9D theory has
a 7-brane and a 6-brane.
If so, these need to be understood in terms of the
M theory. The 6-brane does not raise any new issues, since it is already
present in the 10D IIA theory. It does, however, reinforce our
confidence in the existence of the IIB 7-brane.
A 9D 7-brane, on the other hand, certainly
would require something new in the M theory. What could it be?
To get a 7-brane after compactification on a torus requires
either a 7-brane, an 8-brane, or a 9-brane in the 11D M theory.
However, the cases of $p=7$ and $p=8$ can be ruled out immediately.
They require the existence of a massless vector or scalar particle,
respectively, in the 11D spectrum, and neither of these is present.
The 9-brane, on the other hand,
would couple to a 10-form potential with an 11-form field strength,
which does not describe a propagating mode and therefore cannot be
so easily excluded. Let us therefore
consider the possibility that such a 9-brane with tension
$T_9^{(M)}$ really exists and trace through its consequences in
the same spirit as the preceding discussions.

First we match the 7-brane obtained by wrapping the
hypothetical 9-brane of the
M theory on the spatial torus to the 7-brane obtained from the IIB theory.
This gives the relation
\begin{equation}
A_M T_9^{(M)} = \beta^8 T_{7}^{(B)}. \label{amt9}
\end{equation}
Substituting eq. (\ref{betaeqn}) gives
\begin{equation}
T_7^{(B)}
= \big(A_M\big)^{-1} { T_1^{(B)}T_{9}^{(M)} \over \Big( T_2^{(M)} \Big)^4}.
\label{t7b}
\end{equation}
This formula is not consistent with our assumptions.
A consistent picture would
require $T_7^{(B)}$ to be independent of $A_M$ or $L_B$,
but we have found that $T_7^{(B)} \sim A_M^{-1} \sim L_B^{4/3}$.
Also, the 8-brane and 9-brane of the IIA theory implied by a 9-brane
in the M theory do not have the expected properties.
I'm not certain what to make of all this, but it is tempting to
conclude that there
is no $9$-brane in the M theory. Then, to avoid a paradox
for 9D 7-branes, we must argue that they are not actually present.
I suspect that the usual methods for
obtaining BPS saturated $p$-branes in $d-1$ dimensions from periodic
arrays of them in $d$ dimensions break down for $p = d-3$, because the fields
are not sufficiently controlled at infinity, and therefore
there is no 7-brane in nine dimensions. Another reason to be
suspicious of a 9D 7-brane is that a $(d-2)$-brane in $d$ dimensions
is generically associated with a cosmological term, but straightforward
compactification of the IIB theory on a circle does not give one.

In a recent paper \cite{polchinski},
Polchinski has argued for
the existence of a 9-brane in the 10D IIB theory and an 8-brane
in the 10D IIA theory, both of which carry RR charges.
(He also did a lot of other interesting things.) It ought to be
possible to explore whether the existence of these
objects is compatible with the reasoning of this paper,
but it is unclear to me what the appropriate rules are for handling
such objects.

\section{Conclusion}

We have shown that by
assuming the existence of a quantum `M theory' in eleven dimensions
one can derive a number of non-trivial
relations among various perturbative and non-perturbative structures
of string theory.
Specifically, we have investigated what can
be learned from identifying M theory on $\RR^9 \times T^2$ with type IIB
superstring theory on $\RR^9 \times S^1$ and matching (BPS saturated)
$p$-branes in nine dimensions.
Similarly, we identified the M theory on $\RR^{10} \times S^1$ with type IIA
superstring theory on $\RR^{10}$ and matched $p$-branes in ten dimensions.
Even though quantum M theory surely has no perturbative
definition in 11D Minkowski space, these results
make it more plausible that a non-perturbative quantum theory does exist.
Of course, this viewpoint has
been advocated by others -- most notably Duff and Townsend -- for many years.

Clearly, it would be interesting to explore other
identifications like the ones described here. The natural candidate
to consider next, which is expected to work in a relatively straightforward
way,
is a comparison of the M theory on $\RR^7 \times K3$
with the heterotic string theory
on $\RR^7 \times T^3$. There is a rich variety of $p$-branes
that need to be matched in seven dimensions.
In particular, the M theory 5-brane wrapped on the $K3$ surface
should be identified with the heterotic string itself.

The M theory on $\RR^4 \times S^1 \times K$, where $K$ is a Calabi--Yau space,
should be equivalent to the type IIA superstring
theory on $\RR^4 \times K$. Kachru and
Vafa have discussed examples for which there is a good candidate
for a dual description based on
the heterotic string theory on $\RR^4 \times K3 \times T^2$ \cite {kachru}.
A new element, not encountered in the previous examples, is that
while there is plausibly a connected moduli space of $N=2$ models
that is probed in this way, only part of it is accessed from the
M theory viewpoint and a different (but overlapping) part
from the heterotic string theory viewpoint. Perhaps
this means that we still need to find a theory
that is more fundamental than either the heterotic string theory or
the putative M theory.

\section{Acknowledgments}

I wish to acknowledge discussions with
R. Leigh, N. Seiberg, S. Shenker, L. Susskind, and E. Witten.
I also wish to thank the Rutgers string theory group for its hospitality.
This work was supported in part
by the U.S. Dept. of Energy under Grant No. DE-FG03-92-ER40701.

\end{document}